\documentclass[lettersize,journal]{IEEEtran}
\usepackage{hyperref}
\pdfoutput=1
\usepackage{amsmath,amsfonts}
\usepackage{algorithmic}
\usepackage{array}
\usepackage[caption=false,font=normalsize,labelfont=sf,textfont=sf]{subfig}
\usepackage{textcomp}
\usepackage{stfloats}
\usepackage{url}
\usepackage{verbatim}
\usepackage{graphicx}
\usepackage{cite}
\usepackage{multirow}
\usepackage{siunitx}
\def\BibTeX{{\rm B\kern-.05em{\sc i\kern-.025em b}\kern-.08em
    T\kern-.1667em\lower.7ex\hbox{E}\kern-.125emX}}
\usepackage{balance}
\usepackage[inkscapeformat=png]{svg}
\newcommand{\PMADA}{\mbox{PowerModelsADA}}
\newcommand{\PM}{\mbox{PowerModels}}

\begin{document}

\title{PowerModelsADA: A Framework for Solving Optimal Power Flow using Distributed Algorithms}

\author{Mohannad Alkhraijah,
Rachel Harris,
Carleton Coffrin,
and Daniel K. Molzahn\thanks{Support from NSF AI Institute for Advances in Optimization, \#2112533.}
}

\maketitle

\begin{abstract}
This letter presents \PMADA, an open-source framework for solving Optimal Power Flow (OPF) problems using Alternating Distributed Algorithms (ADA). \PMADA{} provides a framework to test, verify, and benchmark both existing and new ADAs. This letter demonstrates use cases for \PMADA{} and validates its implementation with multiple OPF formulations.
\end{abstract}


\section{Introduction}
\IEEEPARstart{A}{lternating} Distributed Algorithms (ADA) decompose large optimization problems into smaller subproblems to share calculations among multiple computing agents~\cite{molzahn2017survey, ALKHRAIJAH2022108297}. ADAs have potential advantages in scalability, reliability, and communication requirements. These advantages motivate using ADAs to solve power system optimization problems~\cite{molzahn2017survey}. However, there is no standard framework for benchmarking ADAs, leading to implementation, comparison, and replicability challenges that are compounded by different formulations, data structures, and communication requirements among ADAs.

To address these challenges, we present an open-source Julia package called \PMADA{}\footnote{https://github.com/mkhraijah/PowerModelsADA.jl} (\textbf{Power Models} \textbf{A}lternating \textbf{D}istributed \textbf{A}lgorithms) that provides a framework for solving Optimal Power Flow (OPF) problems using ADAs. \PMADA{} uses the power system optimization package \PM{}~\cite{8442948} and the optimization modeling package JuMP~\cite{DunningHuchetteLubin2017} to solve the subproblems. \PM{} uses JuMP to build an optimization model based on a user-selected OPF formulation, and JuMP then passes this optimization model to a user-selected solver. \PMADA{} adds a layer to these tools that permits selecting among multiple ADAs. 

\begin{figure}
    \centering
    \includegraphics[width=\columnwidth]{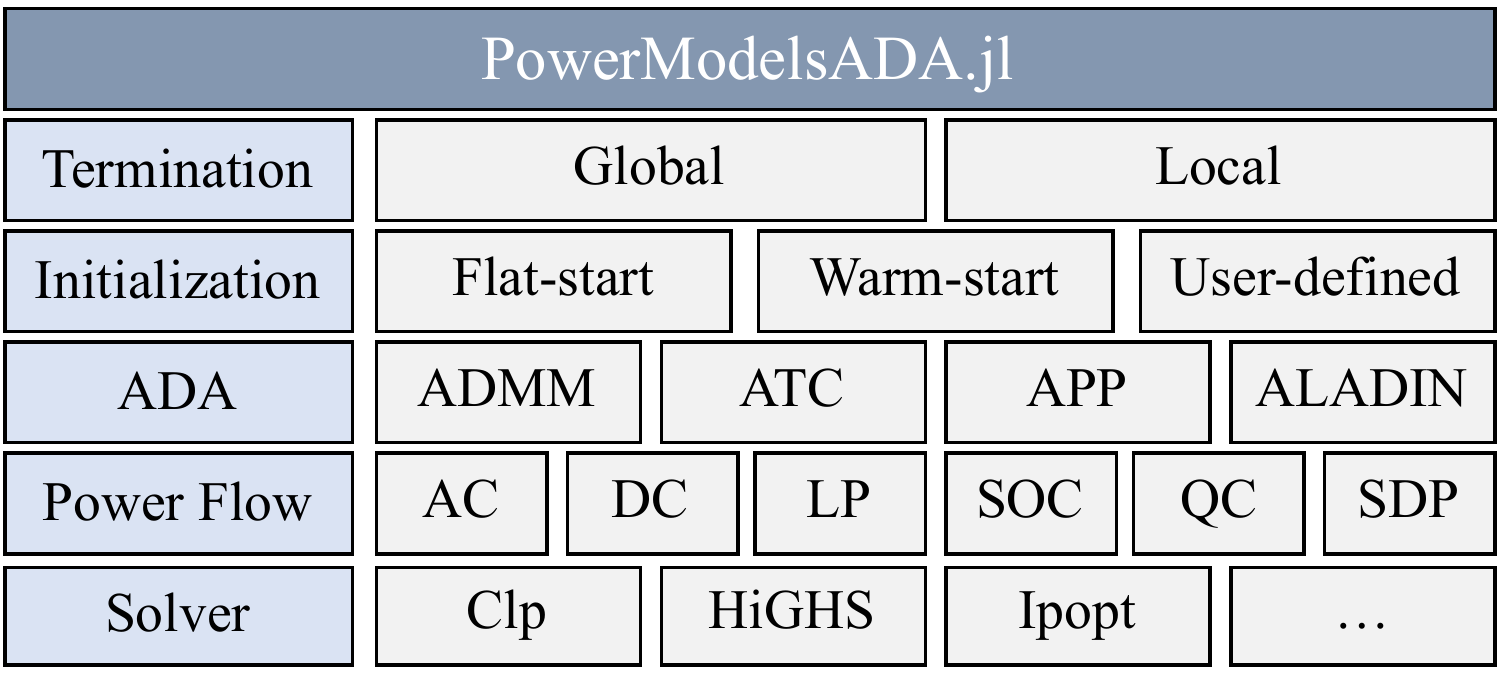}
    \caption{Illustration of \PMADA{} options of termination methods, initializations, ADAs, power flow models, and solvers.}
    \label{fig:PMADA}
\end{figure}

As shown in Fig.~\ref{fig:PMADA}, \PMADA{} enables the ``plug-and-play'' selection among several ADAs, initializations, power flow models, and solvers. With standardized data, communication, and computational structures, \PMADA{} currently implements four ADAs: Alternating Direction Method of Multipliers (ADMM)~\cite{boyd2011distributed}, Analytical Target Cascading (ATC)~\cite{6917065}, Auxiliary Problem Principle (APP)~\cite{kim1997coarse}, and Augmented Lagrangian Alternating Direction Inexact Newton (ALADIN)~\cite{aladin_opf2019}. 

This letter demonstrates the implementation and use cases of \PMADA{}. Section~\ref{sec:OPF} provides the OPF problem formulation and overviews the ADAs. Section~\ref{sec:PMADA} explains the architecture of \PMADA{}. Section~\ref{sec:results} illustrates \PMADA{}'s benchmarking capabilities. Section~\ref{sec:conclusions} presents conclusions and future directions.
\section{Optimal Power Flow} \label{sec:OPF}
\subsection{Problem Formulation}
The OPF problem minimizes an objective, typically generation cost, subject to power flow and engineering constraints. While \PMADA{} implements most of the OPF formulations supported by PowerModels, we present the AC OPF formulation for illustrative purposes:
\begin{subequations}%
\label{eq:acopf}%
\begin{align}%
& \mbox{minimize} \; f_{G} := \sum_{g\in \mathcal{G}} c_{g_2} \Re(S^G_{g})^2 + c_{g_1} \Re(S^G_{g}) + c_{g_0} \label{eq:objective} \\[-0.5em]
& \mbox{subject to: }  \nonumber
\end{align}
\vspace{-20pt}
\begin{align}
&\sum_{\substack{g \in \mathcal{G}_i}}\! S^G_{g}\! -\! \sum_{\substack{l \in \mathcal{L}_i}}\! S^L_{l}\! - \!\sum_{\substack{h \in \mathcal{H}_i}}\!\! (Y^{s}_{h})^* |V_i|^2 \! =\! \sum_{\substack{(i,j)\in \mathcal{E}}}\!\!\!\! S_{ij},\!\!\!\!\!\!\!\!\!\!\!\!\!\!\!\!\!\!\!\!\!\! &\forall i\in \mathcal{N},& \label{eq:power_balance} \\ 
& S_{ij} = Y_{ij}^* V_i V_i^* - Y_{ij}^* V_i V_j^*,  \;\;\qquad\qquad\quad &\forall (i,j)\in \mathcal{E},& \label{eq:power_flow} \\ 
& V_i^{min} \leq |V_i| \leq V_i^{max}, \;\; &\forall i \in \mathcal{N},& \label{eq:voltage_bounds} \\
& S_{g}^{min} \leq S^G_{g} \leq S_{g}^{max}, \;\; &\forall g \in \mathcal{G},& \label{eq:generator_bounds} \\ 
& |S_{ij}| \leq S^{max}_{ij}, \;\; &\forall (i,j) \in \mathcal{E},& \label{eq:branch_bounds} \\ 
& \angle V_{r} = 0,  \;\; &&  \label{eq:references}
\end{align}
\end{subequations}
\noindent where $\mathcal{N}$, $\mathcal{G}$, $\mathcal{L}$, and $\mathcal{H}$ are the sets of buses, generators, loads, and shunts. The set~$\mathcal{E}$ consists of branches connecting two buses with both the forward and backward directions. The subsets $\mathcal{G}_i\subset\mathcal{G}$, $\mathcal{L}_i\subset\mathcal{L}$, and $\mathcal{H}_i\subset\mathcal{H}$ are the corresponding elements at bus $i\in \mathcal{N}$. The decision variables are the buses' voltage phasors $V_i\in\mathbb{C}$, $\forall i \in \mathcal{N}$, the generators' power outputs $S^G_{g}\in\mathbb{C}$, $\forall g \in \mathcal{G}$, and the branches' power flows $S_{ij}\in\mathbb{C}$, $\forall (i,j) \in \mathcal{E}$. The load demands are $S^L_l\in\mathbb{C}$, $\forall l \in \mathcal{L}$, branch admittances are $Y_{ij}\in\mathbb{C}$, $\forall (i,j) \in \mathcal{E}$, and shunt admittances are $Y^{s}_h\in\mathbb{C}$, $\forall h \in \mathcal{H}$. Each generator $g \in \mathcal{G}$ has a quadratic cost function with coefficients $c_{g_2}$, $c_{g_1}$, and $c_{g_0}$. We use $\Re(\,\cdot\,)$, $|\,\cdot\,|$, $\angle (\,\cdot\,)$, and $(\,\cdot\,)^*$ to denote the real part, magnitude, phase angle, and conjugate of complex variables. Complex-valued inequalities are interpreted element-wise on the real and imaginary parts. 

Objective~\eqref{eq:objective} minimizes total generation cost. The equalities~\eqref{eq:power_balance}--\eqref{eq:power_flow} model AC power flow. The inequalities~\eqref{eq:voltage_bounds}--\eqref{eq:branch_bounds} bound voltage magnitudes, generator outputs, and apparent power flows. Constraint~\eqref{eq:references} sets the reference bus~$r$ voltage phase angle.

The formulation in~\eqref{eq:acopf} is commonly called the ACOPF problem since it uses the AC power flow equations. The ACOPF~\eqref{eq:acopf} is non-convex and generally NP-hard~\cite{BIENSTOCK2019494}. Nonetheless, local non-linear solvers can often find good solutions. There are also many OPF approximations and relaxations that are more tractable than~\eqref{eq:acopf}~\cite{molzahn_hiskens-fnt2019}. \PMADA{} builds on \PM's flexibility to select the OPF formulation among the polar (ACP) and rectangular (ACR) forms, various approximations (e.g., DC power flow), and relaxations (e.g., second-order cone (SOC) and quadratic convex (QC))~\cite{8442948}.

\begin{figure}
    \centering
    \includegraphics[width=2.4in]{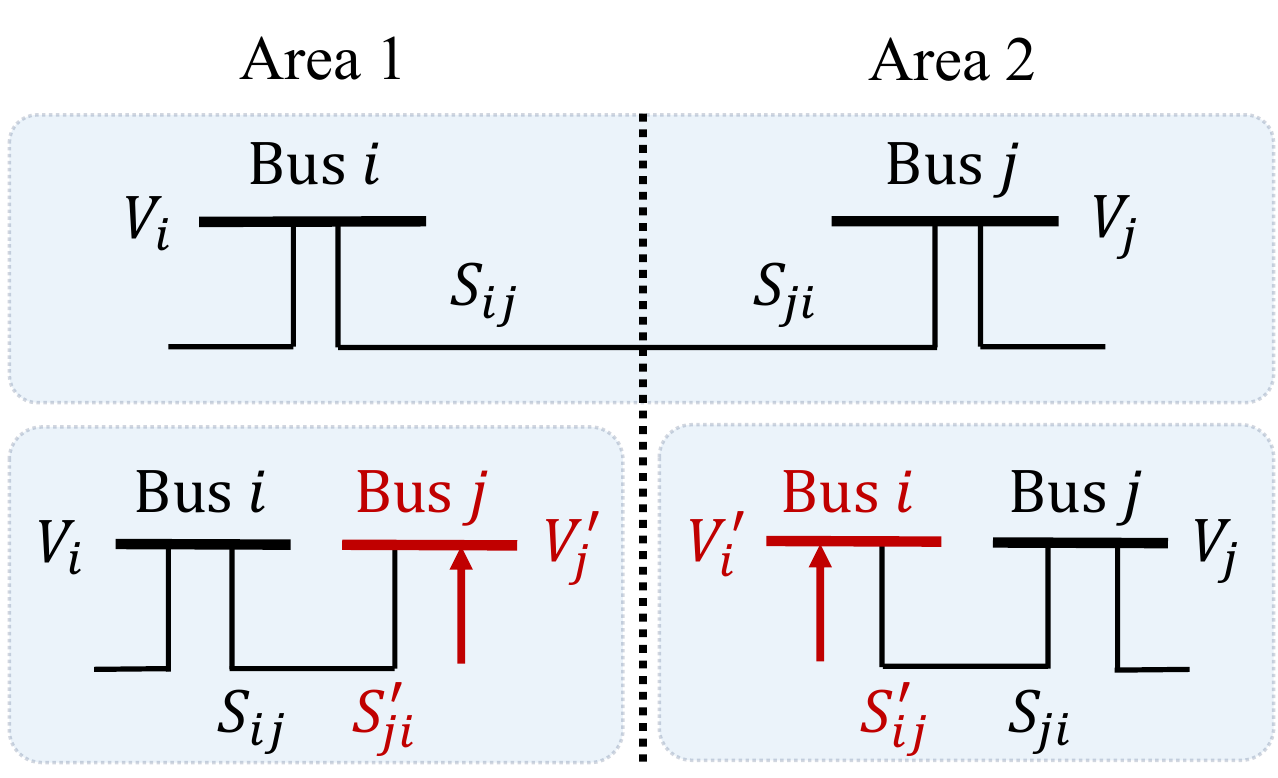}
    \caption{Two connected areas separated by the dashed line before (top) and after (bottom) decomposition.}
    \label{fig:decomposition}
\end{figure}

\subsection{Alternating Distributed Algorithms} \label{sec:DA}
ADAs solve large problems by iteratively solving smaller subproblems associated with each area in the set of areas~$\mathcal{A}$. An area~$a\in\mathcal{A}$ is defined by a set of buses, denoted with $\mathcal{N}^a$, and contains the generators~$\mathcal{G}^a$, loads~$\mathcal{L}^a$, shunts~$\mathcal{H}^a$, and branches~$\mathcal{E}^a$ connected to the buses in $\mathcal{N}^a$. There are several decomposition strategies with differing implications for the number of variables and the convergence rate~\cite{9750783}. \PMADA{} decomposes the OPF problem by introducing fictitious buses and generators at branch terminals connecting two areas as shown in Fig.~\ref{fig:decomposition}. The fictitious generators can inject or absorb arbitrary amounts of active and reactive power with zero-cost unbounded outputs. We then impose consistency constraints between the fictitious and the original variables:
\begin{subequations}%
\label{eq:acopf-modfied}%
\begin{align}%
&&V_{i} &= V^{\prime}_{i}, &\forall i \in \mathcal{N}_B, \\
&&S_{ij} &= S^{\prime}_{ij}, &\forall (i,j) \in \mathcal{E}_B,
\end{align}
\end{subequations}
\noindent where the sets $\mathcal{N}_B$ and $\mathcal{E}_B$ are the boundary buses and branches. To separate the subproblems, we relax the consistency constraints~\eqref{eq:acopf-modfied} using an augmented Lagrangian method with a penalty on the consistency constraints' violations multiplied by a hyperparameter and evaluate the fictitious variables with values shared by neighbors. The ADAs then 1)~solve the subproblems, 2)~share the boundary variable values with neighboring areas, 3)~update the relaxed consistency constraints with the shared boundary variables, and 4)~repeat this process until achieving consensus.

\section{\PMADA{} Architecture} \label{sec:PMADA}
\PMADA{} solves distributed OPF problems via the function \texttt{solve\_dopf}. This function takes the system data, algorithm-specific functions and parameters, the power flow model, and the optimization solver as inputs, and returns the optimal solution of each area. The initial release of \PMADA{} implements four ADAs: ADMM, ATC, APP, and ALADIN.
\begin{figure}[t]
    \centering
    \includegraphics[width=0.85\linewidth]{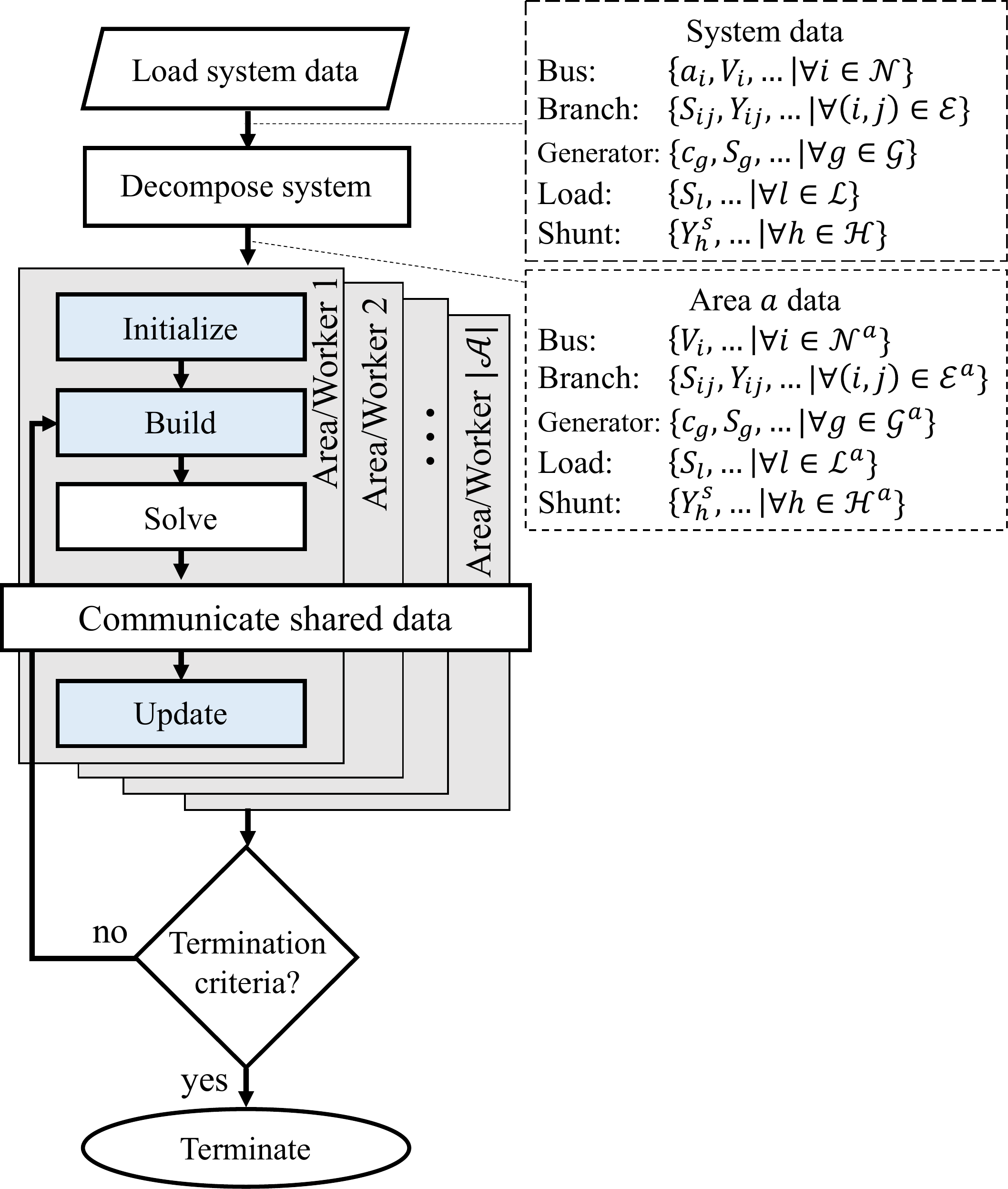}
    \caption{\PMADA{} implementation flowchart. The boxes on the right show the data passed to the next block. The stacked Area/Worker blocks run in parallel. The light-blue blocks indicate algorithm-specific blocks.}
    \label{fig:flowchart}
\end{figure}

Fig.~\ref{fig:flowchart} shows the \PMADA{} algorithmic flowchart. The first two blocks denote the loading and decomposition of system data into multiple areas. The next blocks comprise the agents' local computation, communication, and termination. The light-blue blocks in Fig.~\ref{fig:flowchart} are algorithm-specific blocks, while the white blocks are common to all ADAs. Implementing new ADAs with a similar algorithmic flow only requires defining the algorithm-specific blocks.

Although the algorithmic flow in Fig.~\ref{fig:flowchart} is used by many ADAs, \PMADA{} also implements another algorithmic flow with a central coordinator. Further, the framework can be extended to consider other algorithmic flows (e.g., multiple hierarchical levels) by defining new solve functions. Thus, \PMADA{} facilitates benchmarking both existing and new ADAs by defining a solve function and algorithm-specific blocks. Next, we explain the \PMADA{} data loading, local computation, communication, and termination criteria.

\subsection{Data Loading and Decomposition}
\PMADA{} inherits \PM' ability to load data in M{\sc atpower}~\cite{5491276} and PTI formats, and includes a decomposition function to separate the system data into multiple areas based on the assigned area for each bus as described in Section~\ref{sec:DA}. \PMADA{} uses dictionaries of key-value pairs for data structures, both internally and to interface with \PM{}. The inputs and outputs of the next blocks use the same data structure.

\subsection{Local Computation and Communication} \label{sec:initialization}
The agents receive the area data structure, perform the local computations in parallel using the Distributed library in Julia, and synchronize by communicating their results until achieving the termination criteria. Local computations consist of four functions: initialize, build, solve, and update for the local subproblem. 

\emph{Initialize} is an algorithm-specific function that defines the ADA's shared variables, iteration counter, and status (mismatches and termination criteria) as well as the shared variables' initial values via a flat-start, a warm-start, or a user-defined method. In each iteration, the agents \emph{build} their local subproblems by defining the variables, constraints, and objective. Most ADAs have the same variables and constraints, while the objective is algorithm-specific. The agents then \emph{solve} their subproblem using \PM{} and store the results in the area data structure. Each agent then exchanges shared data with the neighboring areas in the communication block and stores the received data. Then, the agents \emph{update} their area data structure for the next iteration.

\subsection{Termination}
The algorithm terminates after reaching a maximum number of iterations or achieving shared variable mismatches below a specified tolerance, measured via either an $\ell_2$- or $\ell_{\infty}$-norm. Upon termination, the agents store the results in the area data structure. \PMADA{} can check the termination criteria using either central or distributed methods. The central method uses a global variable to store the termination status. The distributed method requires additional iterations to allow agents to communicate the termination status with other agents and terminate the local computations simultaneously.

\begin{table*}[b!]
\caption{{ACOPF Test Case Results}}
\centering
{\begin{tabular}{|l|l||S[table-format=4.1]|S[table-format=4]|S[table-format=4.1]|S[table-format=4]|S[table-format=4.1]|S[table-format=4]|S[table-format=4.1]|S[table-format=4.0]|} 
\hline
\multirow{2}{*}{\textbf{Test case}} & \multirow{2}{*}{\textbf{$\!\!|\mathcal{A}|$}} & \multicolumn{2}{c|}{\textbf{ADMM}} & \multicolumn{2}{c|}{\textbf{ATC}} & \multicolumn{2}{c|}{\textbf{APP}} & \multicolumn{2}{c|}{\textbf{ALADIN}} \\ \cline{3-10} 
  &   & \textbf{Time} & \textbf{Itr.} & \textbf{Time} & \textbf{Itr.} &\textbf{Time} & \textbf{Itr.} & \textbf{Time} & \textbf{Itr.} \\ \hline\hline
14\_ieee   & 2  & 2.1     & 28   & 2.5     & 28   & 2.4     & 31   & 2.0    & 9  \\ \hline
24\_rts    & 4  & 9.3     & 97   & 6.7     & 68   & 20.0    & 207  & 7.1    & 16 \\ \hline
30\_ieee   & 3  & 2.2     & 24   & 3.1     & 33   & 2.4     & 25   & 3.2    & 10 \\ \hline
30\_pwl    & 3  & 2.5     & 24   & 3.9     & 36   & 5.1     & 49   & NA     & NA \\ \hline
39\_epri   & 3  & 9.9     & 89   & 103.0   & 761  & 101.7   & 873  & 8.4    & 20 \\ \hline
73\_rts    & 3  & 10.5    & 61   & 10.5    & 58   & 14.8    & 83   & 73.9   & 35 \\ \hline
179\_goc   & 3  & 27.0    & 66   & 16.9    & 41   & 64.6    & 166  & NC     & NC \\ \hline
300\_ieee  & 4  & 21.4    & 65   & 30.6    & 85   & 28.5    & 77   & 1255.2 & 60 \\ \hline
588\_sdet  & 8  & 301.8   & 870  & 441.3   & 1136 & 462.4   & 1277 & NC     & NC \\ \hline
2000\_goc  & 3  & 3956.2  & 671  & 9374.3  & 1353 & 5863.2  & 1006 & NC     & NC \\ \hline
2853\_sdet & 38 & 8293.8  & 5737 & 3973.2  & 2674 & 4901.4  & 3384 & NC     & NC \\ \hline
4661\_sdet & 22 & 13193.1 & 2951 & 14170.4 & 2662 & 17245.4 & 3747 & NC     & NC \\ \hline
\end{tabular}}
\label{tab:results_ac}
\end{table*}

\section{Test Case and Benchmarking} \label{sec:results}

This section presents two use cases demonstrating the capabilities of \PMADA{}. We first benchmarked four ADAs solving ACOPF problems~\eqref{eq:acopf} with 12 test systems from PGLib-OPF~\cite{PGLib}. We then solved OPF problems using the ADMM algorithm with five power flow formulations for the 588-bus system with eight areas from PGLib-OPF~\cite{PGLib}. See the \PMADA{} repository~\cite{PMADA} for the results from other ADAs with additional test cases. The results here were created with \PMADA~v0.3 in Julia~v1.8 with the Ipopt solver on a high-performance computing platform with 16-cores and 16~GB of memory.

An ADA's performance depends on the choice of hyperparameters that can be challenging to tune. We tuned the hyperparameters of the ADAs by starting with a large value ($10^6$ for ADMM and APP, and $1.2$ for ATC) and then reducing the hyperparameters gradually (dividing by $10$ for ADMM and APP, and subtracting $0.05$ for ATC) until we found a good setting. For ALADIN, we selected the hyperparameters by iterating through a range of values. We used the central termination method with an $l_2$-norm of the mismatch less than $0.01$ (radians and per unit), and reported the results that achieved an objective function value within $1\%$ of the solution from \PM.

For the ADAs benchmarked with the ACOPF formulation~\eqref{eq:acopf}, \PMADA{} produces the results in Table~\ref{tab:results_ac}. The columns $|\mathcal{A}|$ and ``Itr.'' denote the number of areas and iterations. Wall-clock computation times are given in seconds without including the data loading and code precompilation time. ALADIN failed to converge for the six cases marked with ``NC'', likely due to the inherent difficulty in finding appropriate values for this algorithm's ten hyperparameters. Also, ALADIN is inapplicable to ``30\_pwl'' due to non-differentiability of the generators' piecewise-linear cost functions. For the converged test cases, ALADIN has the lowest number of iterations but not necessarily the lowest computation time. Comparing across the ADAs, none of the algorithms outperform the others in all test cases.

\begin{figure}
    \centering
    \includegraphics[width=\columnwidth]{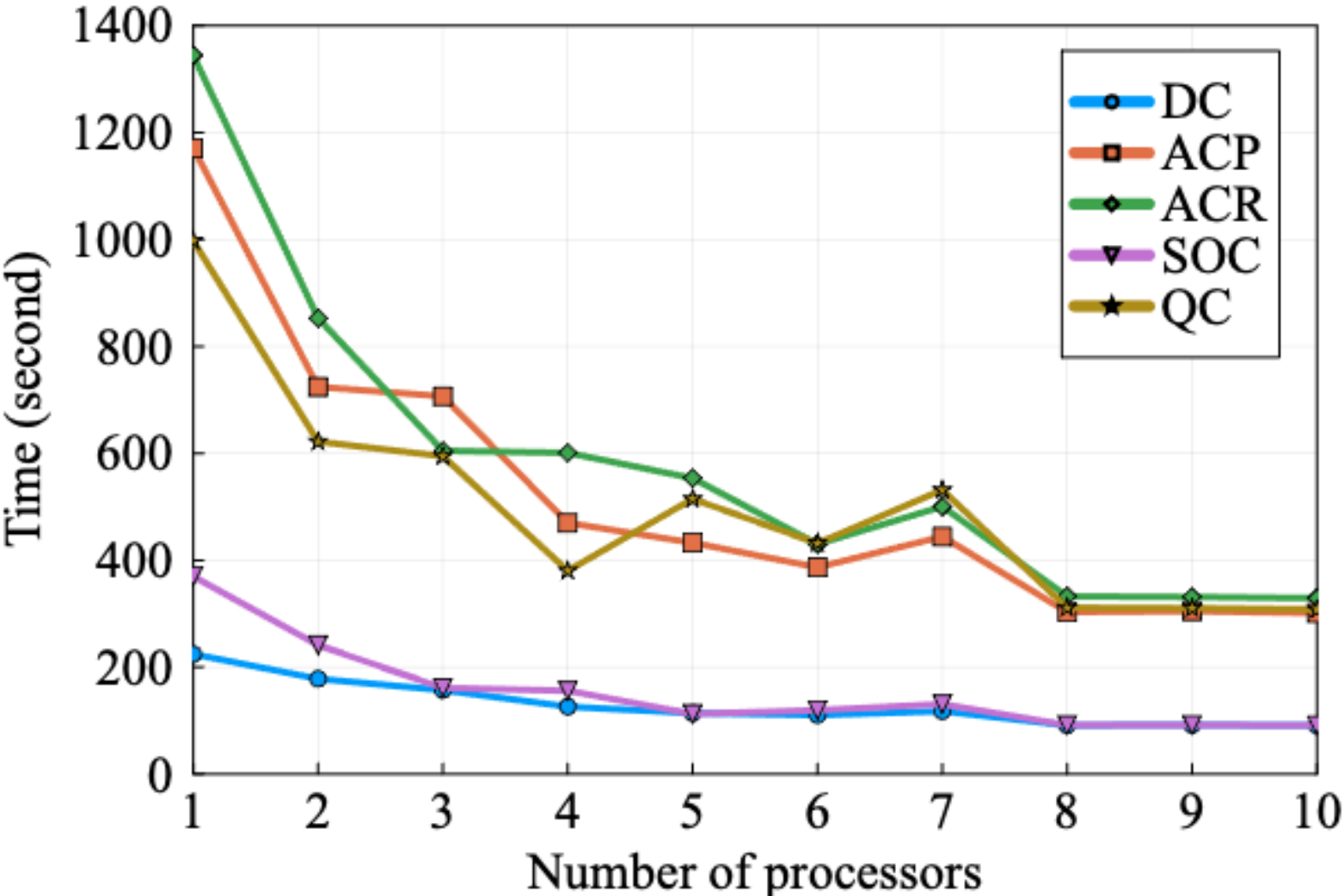}
    \caption{Computation time of solving OPF using five power flow formulations with various number of processors.}
    \label{fig:result}
\end{figure}

We then solved OPF problems with five power flow formulations using the ADMM algorithm for the 588-bus system with eight areas. We repeatedly ran the test case with varying numbers of parallel computations from 1 to 10 processors. We set the value of the hyperparameter to $10^5$ and calculated the computation time by taking the average time of five runs. The computation time of solving the OPF problem with five power flow formulations while varying the number of processors is shown in Fig.~\ref{fig:result}. As we increase the number of processors, the computation time decreases until reaching eight processors. In some cases, the computation time increases with more processors due to the uneven assignment of the subproblems to the processors. The computation time reduces by a factor of four when using eight processors compared to a single processor. Increasing the number of processors beyond eight does not reduce the computation time because the test case has eight areas/workers such that the additional processors are not used. Comparing the power flow formulations, the DC approximation and the SOC relaxation have the fastest convergence rate.

\section{Conclusion and Future Work} \label{sec:conclusions}
\PMADA{} provides a plug-and-play modular framework to test and benchmark ADAs for solving OPF problems. The framework makes it straightforward to implement newly developed ADAs by defining three blocks of code (initialization, building, and updating functions). Users can then solve the OPF problem using ADAs with multiple power flow formulations and optimization solvers. Facilitating the implementation of ADAs provides many advantages to the research community for development and benchmarking.

We are pursuing several directions for extending \PMADA. We intend to consider other power system optimization problems besides OPF and complete the functionality that is in \PM. Other possible extensions include incorporating additional decomposition and initialization methods as well as adding more ADAs.

\newpage 
\bibliographystyle{IEEEtran}
\bibliography{IEEEabrv, manuscript.bib}

\begin{thebibliography}{10}
\providecommand{\url}[1]{#1}
\csname url@samestyle\endcsname
\providecommand{\newblock}{\relax}
\providecommand{\bibinfo}[2]{#2}
\providecommand{\BIBentrySTDinterwordspacing}{\spaceskip=0pt\relax}
\providecommand{\BIBentryALTinterwordstretchfactor}{4}
\providecommand{\BIBentryALTinterwordspacing}{\spaceskip=\fontdimen2\font plus
\BIBentryALTinterwordstretchfactor\fontdimen3\font minus
  \fontdimen4\font\relax}
\providecommand{\BIBforeignlanguage}[2]{{%
\expandafter\ifx\csname l@#1\endcsname\relax
\typeout{** WARNING: IEEEtran.bst: No hyphenation pattern has been}%
\typeout{** loaded for the language `#1'. Using the pattern for}%
\typeout{** the default language instead.}%
\else
\language=\csname l@#1\endcsname
\fi
#2}}
\providecommand{\BIBdecl}{\relax}
\BIBdecl

\bibitem{molzahn2017survey}
D.~K. Molzahn \emph{et~al.}, ``A survey of distributed optimization and control
  algorithms for electric power systems,'' \emph{IEEE Trans. Smart Grid},
  vol.~8, no.~6, pp. 2941--2962, 2017.

\bibitem{ALKHRAIJAH2022108297}
M.~Alkhraijah, C.~Menendez, and D.~K. Molzahn, ``Assessing the impacts of
  nonideal communications on distributed optimal power flow algorithms,''
  \emph{Electric Power Syst. Res.}, vol. 212, p. 108297, 2022, {\rm presented
  at} \emph{22nd Power Syst. Comput. Conf. (PSCC 2022)}.

\bibitem{8442948}
C.~Coffrin, R.~Bent, K.~Sundar, Y.~Ng, and M.~Lubin, ``{PowerModels.jl}: An
  open-source framework for exploring power flow formulations,'' in \emph{20th
  Power Syst. Comput. Conf. (PSCC)}, 2018.

\bibitem{DunningHuchetteLubin2017}
I.~Dunning, J.~Huchette, and M.~Lubin, ``{JuMP}: A modeling language for
  mathematical optimization,'' \emph{SIAM Rev.}, vol.~59, no.~2, 2017.

\bibitem{boyd2011distributed}
S.~Boyd, N.~Parikh, E.~Chu, B.~Peleato, and J.~Eckstein, ``Distributed
  optimization and statistical learning via the alternating direction method of
  multipliers,'' \emph{Found. Trends Mach. Learn.}, vol.~3, no.~1, 2010.

\bibitem{6917065}
A.~Kargarian, Y.~Fu, and Z.~Li, ``Distributed security-constrained unit
  commitment for large-scale power systems,'' \emph{IEEE Trans. Power Syst.},
  vol.~30, no.~4, pp. 1925--1936, 2015.

\bibitem{kim1997coarse}
B.~Kim and R.~Baldick, ``Coarse-grained distributed optimal power flow,''
  \emph{IEEE Trans. Power Syst.}, vol.~12, no.~2, pp. 932--939, 1997.

\bibitem{aladin_opf2019}
A.~Engelmann, Y.~Jiang, T.~Muhlpfordt, B.~Houska, and T.~Faulwasser, ``Toward
  distributed {OPF} using {ALADIN},'' \emph{IEEE Trans. Power Syst.}, vol.~34,
  no.~1, pp. 584--594, 2019.

\bibitem{BIENSTOCK2019494}
D.~Bienstock and A.~Verma, ``Strong {NP}-hardness of {AC} power flows
  feasibility,'' \emph{Oper. Res. Lett.}, vol.~47, no.~6, pp. 494--501, 2019.

\bibitem{molzahn_hiskens-fnt2019}
D.~K. Molzahn and I.~A. Hiskens, ``A survey of relaxations and approximations
  of the power flow equations,'' \emph{Found. Trends Electric Energy Syst.},
  vol.~4, no. 1-2, pp. 1--221, 2019.

\bibitem{9750783}
R.~Harris, M.~Alkhraijah, D.~Huggins, and D.~K. Molzahn, ``On the impacts of
  different consistency constraint formulations for distributed optimal power
  flow,'' in \emph{IEEE Texas Power Energy Conf. (TPEC)}, 2022.

\bibitem{5491276}
R.~D. Zimmerman, C.~E. Murillo-Sánchez, and R.~J. Thomas, ``Matpower:
  Steady-state operations, planning, and analysis tools for power systems
  research and education,'' \emph{IEEE Trans. Power Syst.}, vol.~26, no.~1, pp.
  12--19, 2011.

\bibitem{PGLib}
S.~Babaeinejadsarookolaee \emph{et~al.}, ``The power grid library for
  benchmarking {AC} optimal power flow algorithms,'' \emph{arXiv:1908.02788},
  2019.

\bibitem{PMADA}
M.~Alkhraijah, R.~Harris, C.~Coffrin, and D.~K. Molzahn,
  ``{\mbox{PowerModelsADA.jl}},''
  \url{https://github.com/mkhraijah/PowerModelsADA.jl}, 2022.

\end{thebibliography}

LA-UR-23-30591

\end{document}